\documentclass[aps,twocolumn,groupedaddress,nofootinbib,floatfix]{revtex4-1}

\usepackage[pdftex]{graphicx}
\usepackage{hyperref}
\usepackage{color}
\usepackage[english]{babel}
\usepackage[final]{pdfpages}
\usepackage{pgffor}

\newcommand{\be}{\begin{equation}}
\newcommand{\ee}{\end{equation}}

\begin{document}

\title{Fundamental difference
 between superblockers and superspreaders in networks}

\author{Filippo Radicchi}
\affiliation{Center for Complex Networks and Systems Research, School of Informatics and Computing, Indiana University, Bloomington, USA}

\author{Claudio Castellano}
\affiliation{Istituto dei Sistemi Complessi (ISC-CNR), Via dei Taurini 19, 00185 Roma, Italy,\\ and
Dipartimento di Fisica, Sapienza Universit\`a di Roma, Roma, Italy}
\email{claudio.castellano@roma1.infn.it}

\date{\today}

\begin{abstract}
Two important problems regarding spreading phenomena in complex
topologies are the optimal selection of node sets either to minimize or
maximize the extent of outbreaks. Both problems are nontrivial when
a small fraction of the nodes in the network can be used to achieve
the desired goal.  The minimization problem is equivalent to a
structural optimization. The ``superblockers'',
i.e., the nodes that should be removed from the network to minimize the
size of outbreaks, are those nodes that make connected components as small as
possible. ``Superspreaders'' are instead the nodes such that, if
chosen as initiators, they maximize the average size of outbreaks.
The identity of superspreaders is expected to depend not just on the
topology, but also on the specific dynamics considered.
Recently, it has been conjectured that the two optimization problems 
might be equivalent, in the sense that superblockers act also as
superspreaders. In spite of its potential groundbreaking importance,
no empirical study has been performed to validate this conjecture.
In this paper, we perform an extensive analysis over a large set of
real-world networks to test the similarity between sets of
superblockers and of superspreaders. We show that the 
two optimization
problems are not equivalent: superblockers do not act as optimal spreaders.
\end{abstract}


\maketitle

\section{Introduction}

The interplay between structure and function is at the heart
of the interest attracted by the study of complex 
networks in recent years. Processes mediated by disordered
interaction patterns are affected by the topological properties
of the underlying graph in nontrivial 
ways~\cite{Boccaletti2006,newman2010networks,Lu16}.
Spreading phenomena are among the most fundamental
and studied types of dynamics 
occurring on networks~\cite{PastorSatorras15}.
In this context, a natural question, with implications 
for practical applications, is the following: given a network and 
a spreading dynamics on top of it, how can we
identify the set of $n$ ``superspreaders,'' i.e., the
$n$ vertices such that, if the spreading process is initiated
simultaneously by all of them, the average number 
of nodes reached by the spreading event is maximal?
This problem is often indicated also as ``influence maximization,''
in particular in computer science, where fundamental results
have been derived~\cite{Domingos01,Kempe03}.

An equally interesting and important problem is the 
identification of the set of $n$ ``superblockers'' i.e., 
the $n$ vertices such that, if immunized, and thus effectively 
removed from the network, lead to the minimal average size of 
the outbreak.
As spreading may occur only if contacts are present, the identification of
superblockers is equivalent to the solution of the so-called
optimal percolation problem~\cite{Morone15}, i.e., the identification
of the minimum set of nodes to be eliminated in order to destroy the 
giant component of the network.
Superblockers effectively correspond to the nodes that, when removed, 
minimize the size of the largest connected component in the network.
Solving the optimal percolation problem is nontrivial, and 
many interesting results have appeared in the
last few months~\cite{Clusella16,Braunstein16,Mugisha16,Zdeborova16}.

In their seminal work on the problem of optimal percolation~\cite{Morone15},
Morone and Makse
hinted a strong connection between the identification problems
of superspreaders and superblockers.
The paper effectively describes the problem of identifying superblockers, 
but it always refers to superblockers 
as they were optimal spreaders, suggesting that essentially the two sets 
coincide. 
The analogy between superblockers and superspreaders may 
sound plausible from some point of view:
it is natural to expect that both superblockers and superspreaders 
will be found among the nodes with largest connectivity. 
On the other hand, a conspicuous difference between the two problems
is that optimal percolation depends only on the topological structure,
while influence maximization depends (at least in principle) on the 
type of spreading process considered and on the detailed value of the
parameters describing it. 
In Ref.~\cite{Morone15}, caveats about the distinction between the two problems
are put forward, specifying that the mapping between influence maximization
and optimal percolation is exact only for the Linear Threshold Model with a
very particular choice of the thresholds. A similar approach, based purely on
topological information, has been recently used also for more general
choices of the threshold~\cite{Pei16}.
For different types of spreading models, such as for example those
belonging to the susceptible-infected-removed (SIR) class,
methods relying on the mapping to the optimal percolation problem are 
not a priori granted to work.
Nonetheless, the idea that superspreaders and superblockers are equivalent
in arbitrary spreading models has been rapidly adopted without further
scrutiny~\cite{Liu16,Szolnoki16,Gong16,Wang16,lokhov2016,Ni16,cordasco2015,Guo16,gao2016} (Ref.~[8] being an exception); this calls for a 
deeper and more careful investigation.

In this paper, we perform a critical analysis of the conjectured
coincidence between superblockers and superspreaders when the
spreading process is described by the Independent Cascade Model (ICM),
a very simple dynamics belonging to the same class of 
SIR-like models for epidemic spreading.
By applying algorithms to determine independently sets of optimal blockers 
and sets of optimal spreaders for a large collection of real-world topologies,
we are able to show that in general they are very different. 
Moreover we clarify that the identity of superspreaders strongly depends
on the only parameter of the ICM model:
characterizing optimal spreaders based on purely topological network 
properties (with no reference to the specific spreading dynamics) 
is thus an impossible task.

\section{Blockers, spreaders and the observables considered}

\subsection{Identification of superblockers}

The set of superblockers is defined as the minimal
set of vertices such that their removal leaves no extensive component
in the network.
The identification of the superblockers is equivalent to optimizing a
percolation process~\cite{Morone15}.  After its formalization, several
different heuristic strategies have been introduced to perform this
optimization task.  Morone and Makse~\cite{Morone15} proposed a greedy
algorithm based on a quantity, Collective Influence, to rank vertices
according to their blocking power. Later on, Clusella et
al.~\cite{Clusella16} modified the algorithm for explosive
percolation~\cite{achlioptas2009explosive, d2015anomalous} to identify
optimal blockers. 
Other non-greedy approaches are based on belief-propagation
methods~\cite{Mugisha16,Braunstein16}.  Very recently an iterative
method based on the exploitation of the 2-core structure of the
network has been shown to perform well while being computationally
very efficient~\cite{Zdeborova16}.

In our work, we use the first two methods (Collective Influence 
and Clusella et al.) to identify superblockers. 
In particular, we apply the CI$_3$ method, where the Collective Influence
(CI) of a node is computed by summing over nodes at the frontier
of balls of radius $\ell=3$ (see Ref.~\cite{Morone15} for details). 
We use also a simplified version
of the algorithm by Clusella et al., where node scores 
are not computed iteratively but they are set equal
to their degree. This modification of the algorithm by Clusella et al.
reduces slightly its performance, but it
allows us to treat all networks in the same manner,
without the need to determine additional ad-hoc parameter values
for every specific network.
We remark that the CI and Clusella et al. algorithms provide only
sub-optimal solutions to the problem of identifying superblockers. However, 
the solutions they provide are sufficiently close to the optimum, so 
that we do not expect substantial variations if other, possibly
more effective, algorithms for the identification of superblockers are used.

To be more specific, all algorithms devised to obtain a solution of
the optimal percolation problem identify
the minimal set of superblockers able to destroy the giant component
of the graph. The distinction between the extensive (giant) 
component of a network and subextensive components is clear-cut only
in the limit of infinite size. For finite networks such as those
we consider here,
the distinction is blurred and somehow arbitrary. In practice, in our
study we adopt the same convention put forward by Clusella et 
al.~\cite{Clusella16}, and consider a component to be extensive
if its size is larger than $\sqrt{N}$, where $N$ is the overall
number of network vertices.

In general, the methods for the optimal percolation problem provide a
set of a $n_c^{(x)}$ vertices, being $n_c^{(x)}$ a specific value depending
on the method $x$ considered.
However, the goal of our analysis is to test whether superblockers
are also optimal spreaders for a generic set size $n$.
We will therefore use the methods by Morone and Makse, and by Clusella
et al. to assign nodes with a rank ranging from $1$ to $N$. 
In one case, the  order of the nodes will be established as the inverse order
in which they are added in the algorithm  by Clusella et al. In the other case
instead, it will coincide with the order in which nodes are removed 
from the network according to the Collective Influence score.
We will measure the agreement,
as a function of $\rho=n/N$, between the set $S^{(x)}$ of best 
$n$ superblockers found
by methods devised to optimally destroy a network
and the set $S^{(C)}$ of the best $n$ superspreaders identified by a method
specifically deployed for their identification (see below). 
For reference, we will consider also two alternative rankings: a completely
random one (dubbed ``Random'' in the following) and a ranking based on node 
degrees (with random ordering in the case of tie), denoted as
``Degree.''

\subsection{Spreading dynamics}

As spreading dynamics on networks, we consider the Independent Cascade 
Model (ICM)~\cite{Goldenberg01}, in its simplest, unweighted version.
This model is commonly considered in studies of influence maximization 
by computer scientists.
One starts from a set $S$ (of size $n$) of 
initially activated nodes at time $t=0$.
All other nodes are instead initially set as inactive. 
At each discrete time step $t$, two rules are applied in sequence:
(i) Each activated node $i$ contacts all its
neighbors $j$ and, with an independent probability $p$, tries 
to activate each of those nodes that have been never
activated during previous stages of the dynamics; 
(ii) All nodes that tried to activate their neighbors at step (i)
become inactive, and they cannot be activated again in subsequent
stages of the dynamics.
The process is iterated until no more active nodes are present.
This dynamics is a parallel version of the 
common SIR model~\cite{PastorSatorras15} for epidemic
spreading, with the time to recover fixed deterministically to 1.

\subsection{Identification of superspreaders}

The identification of superspreaders 
(or influence maximization) in networks has 
attracted a huge interest in the last 15 years, since its formalization
by Domingos and Richardson~\cite{Domingos01}.
In a nutshell, the problem is the following: given a network of size 
$N$ and a spreading dynamics on top of it, a set $S$ of initially
active nodes generates a cascade (outbreak) of average size $R(S)$.
We will denote $R(S)$ also as ``spreading power'' of set $S$.
Influence maximization aims at identifying, among all
subsets of size $n$, the subset $S^*$ for which $R(S)$ is maximal.
The seminal paper by Kempe et al.~\cite{Kempe03} has shown that 
the influence maximization problem is computationally hard
(NP-complete). However the same paper provides, for the broad
class of submodular dynamics (including the ICM), 
a greedy algorithm able to find a sub-optimal solution, 
provably within 63\% of the optimum~\cite{Kempe03}. 
More precisely, by adding at each time step to $S$
the node which maximizes the marginal increment of $R$, one is guaranteed
that $R(S) \ge (1-1/e) R(S^*)$, where $e$ is the base of the natural 
logarithm.

Many other works have followed, improving the poor computational
efficiency of Kempe's greedy algorithm, while still preserving
the original performance bounds~\cite{Leskovec07,Chen09}.
Nowadays it is possible  to determine influence maximizers for networks 
with billions of nodes~\cite{Nguyen16}.
More recently, the statistical physics community has started to attack
the problem with its tools and 
concepts~\cite{Altarelli13,Zhang16}.

We are not crucially interested in computational efficiency as networks 
with order $10^4$ nodes are sufficient for our purposes.
Therefore to identify superspreaders in the ICM we use the greedy algorithm
version introduced by Chen et. al~\cite{Chen09}, which is based on
the well-known mapping between SIR dynamics and random
percolation~\cite{Newman02}.
As for the case of superblockers identification, the algorithm we use for
superspreaders identification does not determine the actual optimum,
but just a sub-optimal solution.
Also in this case, we reasonably expect this approximation to have
a very limited impact on the results.
The outcome of Chen's algorithm is a ranking of all network nodes 
with an associated spreading power $R(S^{(C)},\rho)$: this value means that
the set of superspreaders of size $n=\rho N$ is made by all nodes ranked
from $1$ to $n$ and that the average number of nodes reached by
a cascade initiated by them is $R(S^{(C)},\rho)$.

Finally, it is important to remark that we are interested here in 
finding optimal multiple spreaders, i.e., sets of vertices
which maximize the extent of the spreading process when seeded
simultaneously in all of them.
A similar but {\em distinct} problem is the search for optimal single 
spreaders, i.e., the nodes which are most influential when
the process is initiated only in one node~\cite{Kitsak10,Radicchi16}.
The two problems are somehow related, but in a nontrivial manner:
good single influencers may share large parts of their influence zone,
so that seeding the outbreak in all of them at the same time
leads to a cascade only slightly larger than those started by
each of them separately.

\subsection{The observables}

Each identification algorithm of superblockers provides
a different ranking of all nodes in the network. 
For each ranking $x$, by means of ICM numerical simulations
repeated $10,000$ times, we compute 
the spreading power $R(S^{(x)},\rho)$ for any size $\rho N$ of the
seed set $S^{(x)}$.

There are two possible ways to compare the sets of 
superblockers and superspreaders. 

The first possibility is to compare
the identity of the individual nodes. Are the vertices identified
as superblockers also those identified by Chen's algorithm as superspreaders?
To answer this question, we consider the Jaccard index (or similarity)
among the two sets 
$S^{(x)}$ and  $S^{(C)}$. 
This quantity is defined as
\be
J^{(x)}(\rho)  = \frac{|S^{(x)} \cap S^{(C)}|}
{|S^{(x)} \cup S^{(C)}|}
\label{eq1}
\ee
where $|A|$ stands for the number of elements in the set $A$. 
Clearly, if the two sets $S^{(x)}$ and $S^{(C)}$ 
coincide their similarity goes to 1, while it vanishes
if they have null intersection.

The second possibility is to compare not the identity but only the 
spreading power of the two different sets. 
Indeed, it is in principle possible that the set of superblockers does 
not coincide with the set of the best spreaders,
yet it has comparable spreading power. In such a case one would 
conclude that the search for the best blockers effectively uncovers
a set of almost optimal spreaders.
To compare the spreading power of superblockers and of
superspreaders, we consider the ratio $R(S^{(x)},\rho)/R(S^{(C)},\rho)$.
A value of this quantity equal to 1 indicates that the best blockers
are also the best spreaders in the network, while small values 
show that blockers are not good spreaders.

In the evaluation of these comparisons, one must always
keep in mind that in the limit $\rho \to 1$ all sets unavoidably
coincide; hence the quantities defined above tend to 1.

\begin{figure*}
\includegraphics[width=0.9\textwidth]{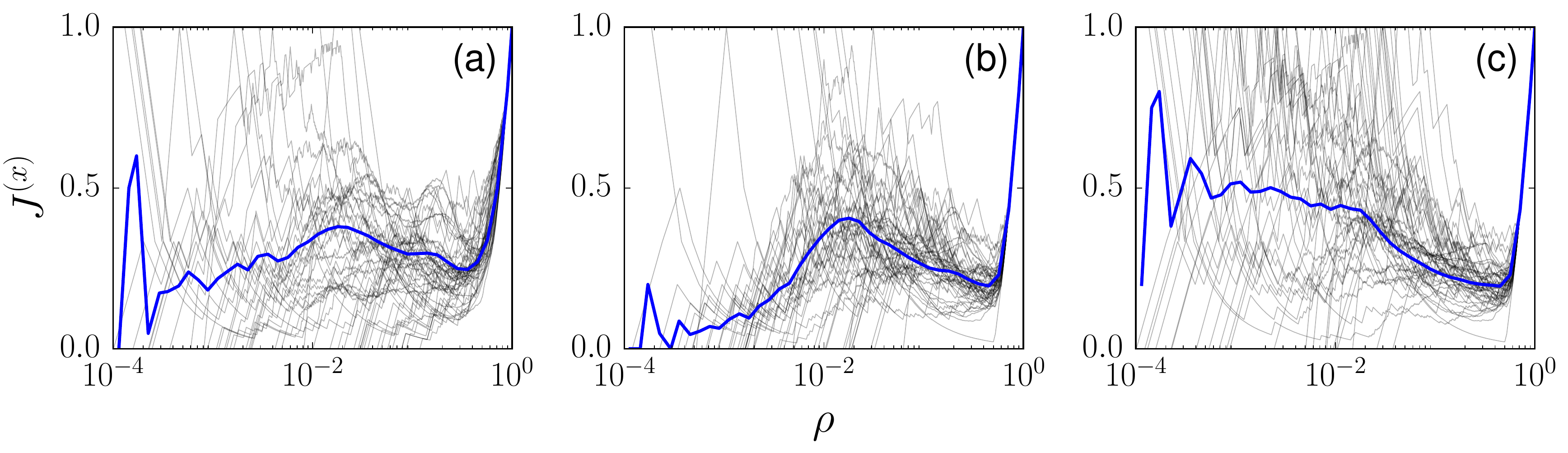}
\caption{\label{fig1} (color online)
Jaccard similarity $J^{(x)}$ between sets of superblockers and of
superspreaders as a function of the fraction $\rho$ of nodes
in the sets. Thin gray lines denote results for each of
the $51$ real-world networks considered in the analysis.
Thick blue lines represent average values of $J^{(x)} (\rho)$
across all networks. Average lines are calculated dividing the range
of possible values of $\rho$ into $50$ equally spaced bins in the
logarithm scale, and computing the average value of $J^{(x)}$ across
all networks within each of those bins.
The set of top $\rho N$ spreaders was identified 
using the method by Chen et al. for the critical value of the
spreading probability $p = p_c$. Superblockers were ranked using  
(a) Collective Influence, (b) Clusella et al., and (c) Degree.
}
\end{figure*}

\subsection{The networks}

As substrate of the optimal percolation and of the ICM spreading 
process we consider $51$ real-world networks 
of very diverse origin, size and topological features~\cite{SM}. 
For each of them we compute the critical value $p_c$ separating the
region of the phase-diagram where outbreaks are subextensive ($p<p_c$)
from the supercritical phase ($p>p_c$) where outbreaks reach
a finite fraction of the whole network. 
The value of $p_c$ is determined as the position of the maximum of the
susceptibility $\langle s^2\rangle/\langle s\rangle^2$ (where
$\langle s^n\rangle$ is the $n$-th moment of the outbreak size 
distribution computed for random initial single 
spreaders)~\cite{Castellano16}. $p_c$ values for each network are 
reported in~\cite{SM}.

\begin{figure}
\includegraphics[width=0.95\columnwidth]{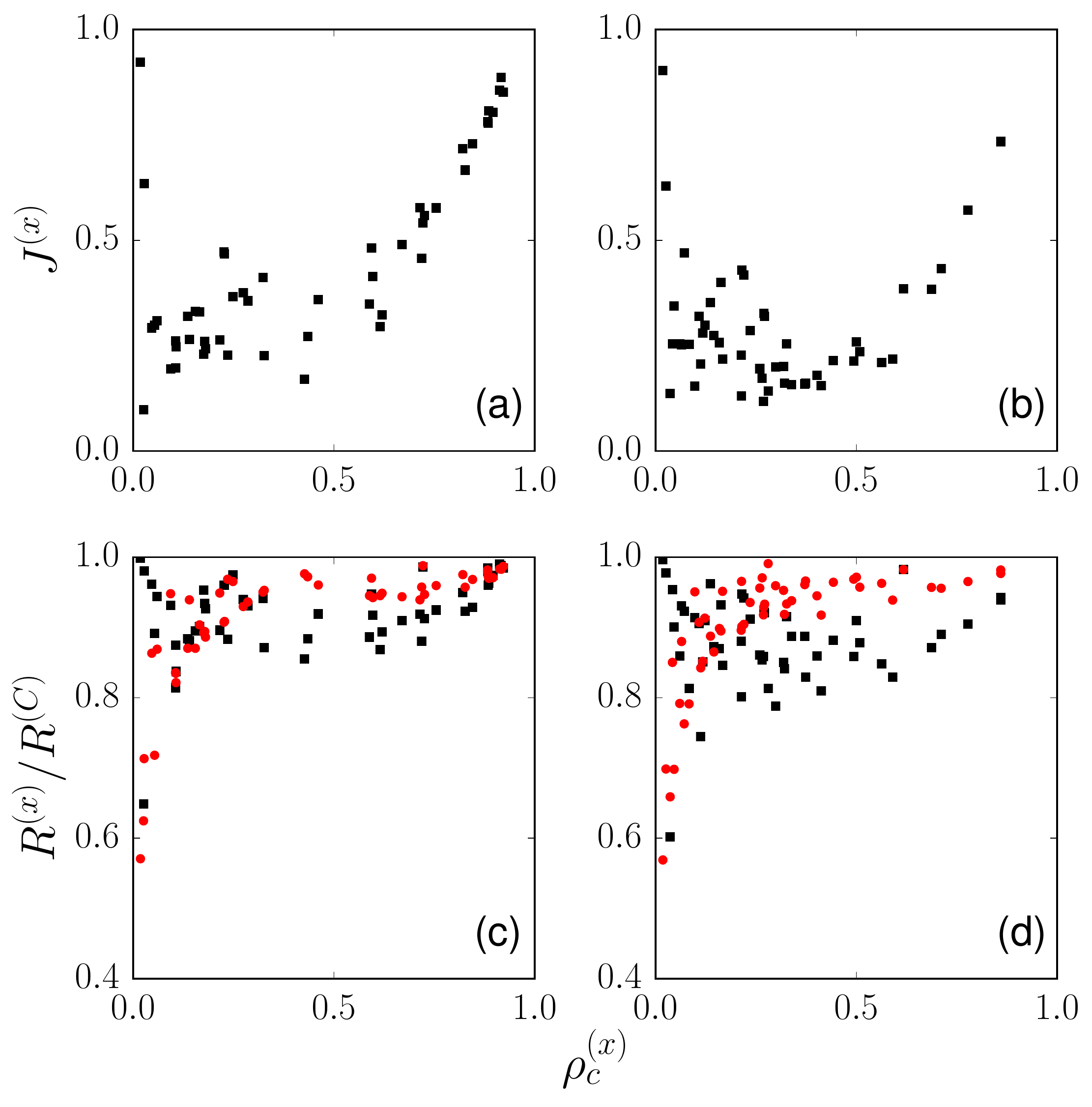}
\caption{\label{fig2} (color online)
(a) Plot of the Jaccard similarity $J^{(x)}$ (see Eq.~(\ref{eq1})) 
for $\rho = \rho_c^{(x)}$. We consider here Collective Influence.
Each point represents a single network.
(b) Same as in panel (a), but for the algorithm by Clusella et al.
(c) Ratio $R^{(x)}/R^{(C)}$ computed at $\rho =
\rho_c^{(x)}$ for the Collective Influence algorithm (black squares).
As a term of comparison, we consider also results of the same
quantity calculated with a Random placement of the same number of
seeds (red circles).
(d) Same as in panel (c), but for the algorithm by 
Clusella et al.
}
\end{figure}

\section{Results}

We start by comparing superblockers with superspreaders
determined when $p=p_c$.

In Fig.~\ref{fig1} we plot, as a function of $\rho$, the value
of the Jaccard similarity $J^{(x)}$ between the set of superspreaders
determined by Chen's greedy algorithm and sets of superblockers
determined using Collective Influence, Clusella et al. method and, as a reference,
the Degree method.
It turns immediately clear that there is a huge variability among
the different cases. However,
by looking at the average value of $J^{(x)}$ (depicted in blue), 
two conclusions can be drawn.
First, there are in general rather few superblockers nodes which belong 
also to the set of superspreaders; 
second, spreading power is more correlated to degree than to the blocking
ability.

One may wonder whether the results of Fig.~\ref{fig1} are due
to the lack of accuracy of the method based on Collective Influence and 
the algorithm by Clusella et al. to establish a precise rank for superblockers.
Both these methods are in fact devised to optimally destroy a network
by finding the minimal set of size $n_c^{(x)} = \rho_c^{(x)} N$ 
whose removal leads to the
disappearance of the giant component in the graph, but these nodes
are not chosen in a special order. We therefore extract from
Fig.~\ref{fig1}
the values of the Jaccard similarity corresponding to $\rho_c^{(x)}$,
and plot them in Fig.~\ref{fig2}a and b. We note that the similarity
between superspreaders and superblockers is still very low, except
for networks with high values of $\rho_c^{(x)}$ (for $\rho_c^{(x)} \to 1$
the two sets obviously tends to coincide).

\begin{figure*}
\includegraphics[width=0.9\textwidth]{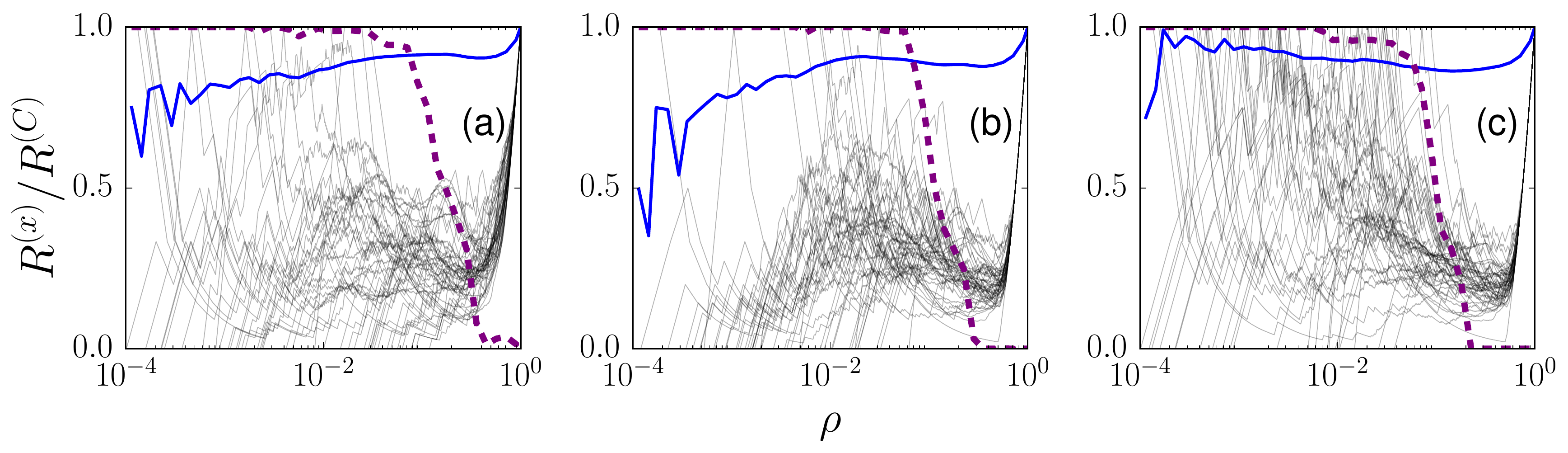}
\caption{\label{fig3} (color online)
  Ratio $R^{(x)}/R^{(C)}$ between the spreading
  power $R^{(x)}$ of the top $\rho N$ superblockers identified by
  criterion $x$ and the spreading power $R^{(C)}$ of the top $\rho N$
  superspreaders identified by the algorithm by Chen et al. The ratio
  $R^{(x)}/R^{(C)}$ is plotted as a function of $\rho$. Thin gray
  lines denote results for each of the $51$ real-world networks
  considered in our analysis.  Thick solid blue lines represent instead
  average values of $R^{(x)}/R^{(C)}$ across all networks. The dashed solid
  purple 
  lines quantify the probability that the actual value of $R^{(x)}$ is
  better than the one obtained by placing the same number of seeds at
  random.  Thick solid blue and think dashed purple lines are calculated 
dividing the range
of possible values of $\rho$ into $50$ equally spaced bins in the
logarithm scale, and computing within each of 
those bins the average value of $J^{(x)}$ across all networks 
(thick solid blue lines) or the frequency of networks for which 
$R^{(x)} \leq R^{(\textrm{\tiny{Random}})}$ (thick dashed purple lines).
Superblockers were identified using (a) Collective
  Influence, (b) Clusella et al., and (c) Degree.  }
\end{figure*}

Figs.~\ref{fig2}a and ~\ref{fig2}b provide strong evidence that sets of 
superspreaders and superblockers are very different.
Nevertheless, one could hypothesize that, even if superblockers
are not the very best spreaders, they still are very good spreaders.
For this reason, in Figs.~\ref{fig2}c and~\ref{fig2}d we plot the ratio 
$R^{(x)}/R^{(C)}$ of the spreading power of blockers to 
the optimal spreading power obtained using Chen's algorithm.
It turns out that the sets of blockers identified using both Clusella
and CI methods are far from being optimal spreaders: their performance
is often even worse than the one resulting by randomly selecting
the same fraction $\rho_c^{(x)}$ of seeds. Similar results are
confirmed in Fig.~\ref{fig3}, where we consider the ratio 
$R^{(x)}/R^{(C)}$ for arbitrary values of $\rho$. 
Superblockers are never good spreaders. 
Ranking nodes based on their degree is generally a much better
strategy than ranking nodes using Collective Influence scores or the
Clusella et al. algorithm. In addition, for sufficiently large values
of $\rho$, generally comparable with $\rho_c^{(x)}$, topological methods
for the identification of superblockers never exceed the performance
of random selection.

The results displayed above are obtained when the
Independent Cascade Model for spreading is at criticality. 
We expect the difference between good and bad spreaders to be maximal
for $p=p_c$. We have repeated the same analysis for other values of $p$,
both well below the critical point ($p=p_c/2$) and well into the supercritical
phase ($p=2 p_c$). The results, reported in~\cite{SM}, 
confirm that in the whole phase-diagram the nodes that keep
the network together (blockers) have no special spreading capability.

\begin{figure*}
\includegraphics[width=0.9\textwidth]{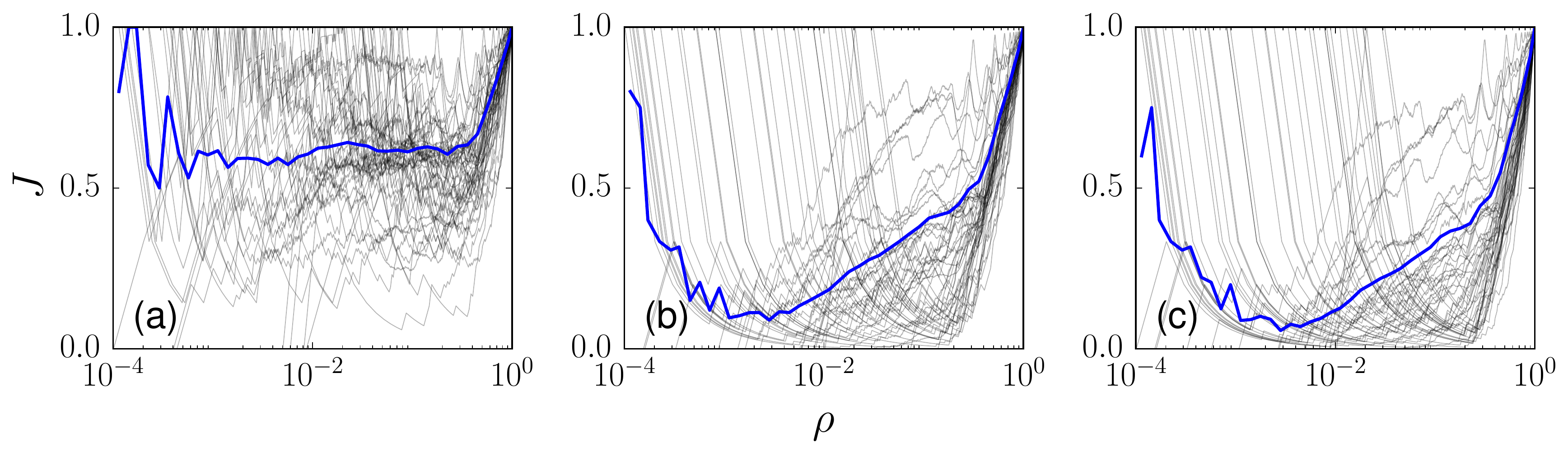}
\caption{\label{fig5} (color online)
Jaccard similarity index as a function of the fraction
$\rho$ of nodes.
(a) Comparison between best spreaders for $p = 0.5 p_c$ and 
$p = p_c$;
(b) comparison between best spreaders for $p = 2 p_c$ and $p = p_c$;
(c) comparison between best spreaders for $p = 2 p_c$ and $p = 0.5
p_c$.
Thin gray lines denote results for each of
the $51$ real-world networks considered in the analysis.
Thick blue lines represent average values of $J(\rho)$
across all networks. Average lines are calculated dividing the range
of possible values of $\rho$ into $50$ equally spaced bins in the
logarithm scale, and computing the average value of $J$ across
all networks within each of those bins.
}
\end{figure*}

The conjecture that superspreaders and superblockers are essentially
the same nodes in a network rests implicitly on the assumption that the 
identity of best spreaders does not depend on the parameter $p$.
We test this hypothesis in Fig.~\ref{fig5}, where we plot, as a function
of $\rho$, the Jaccard distance among sets of optimal spreaders for sub-, 
super- and critical values of $p$. Interestingly, the
sets of optimal spreaders for subcritical and critical evolution 
are quite similar, while they are very different from the optimal set
of spreaders in the supercritical regime.
Whether a set of nodes
has large spreading power crucially depends on $p$. 
Any attempt to
relate sets of optimal spreaders to sets determined only by topology
is ill-fated.

\section{Conclusions}

In this paper we have shown that, for the Independent Cascade Model in 
a network, 
superspreaders and superblockers are two distinct concepts, with
no direct practical connection. More in detail, our results indicate that 
the nodes whose removal leads to the breakdown of the topology into
nonextensive components do {\em not} coincide with the best nodes
for seeding a spreading process. Even the plain degree centrality
identifies better spreaders than the methods aimed at identifying
superblockers. 
In addition, as the identity of the optimal spreaders is 
strongly dependent on the parameter that regulates the dynamics,
attempts to identify sets of superspreaders based
only on topological properties without reference to the details of
the spreading dynamics are bound to fail.

With the benefit of hindsight these results appear rather easy to be
anticipated: the choice of optimal seeds depends
on the spreading dynamics (and its parameters) while optimal blocking
does not. As most recent papers in the field have implicitly assumed 
the validity of this 
conjecture~\cite{Liu16,Szolnoki16,Gong16,Wang16,
lokhov2016,Ni16,cordasco2015,Guo16,gao2016}, we
believe that a detailed verification
was in order.
The minimization and the maximization of the extent of spreading
processes mediated by complex topologies are both exciting examples
of the nontrivial interplay between structure and function.
They are deservedly attracting a huge interest in statistical physics,
computer science and other communities. The interest on these issues
is by no means reduced by the awareness that they are fundamentally
different problems.

\begin{acknowledgments}
FR acknowledges support from the National Science Foundation
(CMMI-1552487) and the US Army Research Office (W911NF-16-1-0104).
\end{acknowledgments}


\begin{thebibliography}{34}%
\makeatletter
\providecommand \@ifxundefined [1]{%
 \@ifx{#1\undefined}
}%
\providecommand \@ifnum [1]{%
 \ifnum #1\expandafter \@firstoftwo
 \else \expandafter \@secondoftwo
 \fi
}%
\providecommand \@ifx [1]{%
 \ifx #1\expandafter \@firstoftwo
 \else \expandafter \@secondoftwo
 \fi
}%
\providecommand \natexlab [1]{#1}%
\providecommand \enquote  [1]{``#1''}%
\providecommand \bibnamefont  [1]{#1}%
\providecommand \bibfnamefont [1]{#1}%
\providecommand \citenamefont [1]{#1}%
\providecommand \href@noop [0]{\@secondoftwo}%
\providecommand \href [0]{\begingroup \@sanitize@url \@href}%
\providecommand \@href[1]{\@@startlink{#1}\@@href}%
\providecommand \@@href[1]{\endgroup#1\@@endlink}%
\providecommand \@sanitize@url [0]{\catcode `\\12\catcode `\$12\catcode
  `\&12\catcode `\#12\catcode `\^12\catcode `\_12\catcode `\%12\relax}%
\providecommand \@@startlink[1]{}%
\providecommand \@@endlink[0]{}%
\providecommand \url  [0]{\begingroup\@sanitize@url \@url }%
\providecommand \@url [1]{\endgroup\@href {#1}{\urlprefix }}%
\providecommand \urlprefix  [0]{URL }%
\providecommand \Eprint [0]{\href }%
\providecommand \doibase [0]{http://dx.doi.org/}%
\providecommand \selectlanguage [0]{\@gobble}%
\providecommand \bibinfo  [0]{\@secondoftwo}%
\providecommand \bibfield  [0]{\@secondoftwo}%
\providecommand \translation [1]{[#1]}%
\providecommand \BibitemOpen [0]{}%
\providecommand \bibitemStop [0]{}%
\providecommand \bibitemNoStop [0]{.\EOS\space}%
\providecommand \EOS [0]{\spacefactor3000\relax}%
\providecommand \BibitemShut  [1]{\csname bibitem#1\endcsname}%
\let\auto@bib@innerbib\@empty
\bibitem [{\citenamefont {Boccaletti}\ \emph {et~al.}(2006)\citenamefont
  {Boccaletti}, \citenamefont {Latora}, \citenamefont {Moreno}, \citenamefont
  {Chavez},\ and\ \citenamefont {Hwang}}]{Boccaletti2006}%
  \BibitemOpen
  \bibfield  {author} {\bibinfo {author} {\bibfnamefont {S.}~\bibnamefont
  {Boccaletti}}, \bibinfo {author} {\bibfnamefont {V.}~\bibnamefont {Latora}},
  \bibinfo {author} {\bibfnamefont {Y.}~\bibnamefont {Moreno}}, \bibinfo
  {author} {\bibfnamefont {M.}~\bibnamefont {Chavez}}, \ and\ \bibinfo {author}
  {\bibfnamefont {D.-U.}\ \bibnamefont {Hwang}},\ }\href {\doibase
  http://dx.doi.org/10.1016/j.physrep.2005.10.009} {\bibfield  {journal}
  {\bibinfo  {journal} {Physics Reports}\ }\textbf {\bibinfo {volume} {424}},\
  \bibinfo {pages} {175 } (\bibinfo {year} {2006})}\BibitemShut {NoStop}%
\bibitem [{\citenamefont {Newman}(2010)}]{newman2010networks}%
  \BibitemOpen
  \bibfield  {author} {\bibinfo {author} {\bibfnamefont {M.}~\bibnamefont
  {Newman}},\ }\href@noop {} {\emph {\bibinfo {title} {Networks: an
  introduction}}}\ (\bibinfo  {publisher} {OUP Oxford},\ \bibinfo {address}
  {New York, NY},\ \bibinfo {year} {2010})\BibitemShut {NoStop}%
\bibitem [{\citenamefont {Lu}\ \emph {et~al.}(2016)\citenamefont {Lu},
  \citenamefont {Chen}, \citenamefont {Ren}, \citenamefont {Zhang},
  \citenamefont {Zhang},\ and\ \citenamefont {Zhou}}]{Lu16}%
  \BibitemOpen
  \bibfield  {author} {\bibinfo {author} {\bibfnamefont {L.}~\bibnamefont
  {Lu}}, \bibinfo {author} {\bibfnamefont {D.}~\bibnamefont {Chen}}, \bibinfo
  {author} {\bibfnamefont {X.-L.}\ \bibnamefont {Ren}}, \bibinfo {author}
  {\bibfnamefont {Q.-M.}\ \bibnamefont {Zhang}}, \bibinfo {author}
  {\bibfnamefont {Y.-C.}\ \bibnamefont {Zhang}}, \ and\ \bibinfo {author}
  {\bibfnamefont {T.}~\bibnamefont {Zhou}},\ }\href {\doibase
  http://dx.doi.org/10.1016/j.physrep.2016.06.007} {\bibfield  {journal}
  {\bibinfo  {journal} {Physics Reports}\ }\textbf {\bibinfo {volume} {650}},\
  \bibinfo {pages} {1 } (\bibinfo {year} {2016})}\BibitemShut {NoStop}%
\bibitem [{\citenamefont {Pastor-Satorras}\ \emph {et~al.}(2015)\citenamefont
  {Pastor-Satorras}, \citenamefont {Castellano}, \citenamefont {Van~Mieghem},\
  and\ \citenamefont {Vespignani}}]{PastorSatorras15}%
  \BibitemOpen
  \bibfield  {author} {\bibinfo {author} {\bibfnamefont {R.}~\bibnamefont
  {Pastor-Satorras}}, \bibinfo {author} {\bibfnamefont {C.}~\bibnamefont
  {Castellano}}, \bibinfo {author} {\bibfnamefont {P.}~\bibnamefont
  {Van~Mieghem}}, \ and\ \bibinfo {author} {\bibfnamefont {A.}~\bibnamefont
  {Vespignani}},\ }\href {\doibase 10.1103/RevModPhys.87.925} {\bibfield
  {journal} {\bibinfo  {journal} {Rev. Mod. Phys.}\ }\textbf {\bibinfo {volume}
  {87}},\ \bibinfo {pages} {925} (\bibinfo {year} {2015})}\BibitemShut
  {NoStop}%
\bibitem [{\citenamefont {Domingos}\ and\ \citenamefont
  {Richardson}(2001)}]{Domingos01}%
  \BibitemOpen
  \bibfield  {author} {\bibinfo {author} {\bibfnamefont {P.}~\bibnamefont
  {Domingos}}\ and\ \bibinfo {author} {\bibfnamefont {M.}~\bibnamefont
  {Richardson}},\ }in\ \href {\doibase 10.1145/502512.502525} {\emph {\bibinfo
  {booktitle} {Proceedings of the Seventh ACM SIGKDD International Conference
  on Knowledge Discovery and Data Mining}}},\ \bibinfo {series and number} {KDD
  '01}\ (\bibinfo  {publisher} {ACM},\ \bibinfo {address} {New York, NY, USA},\
  \bibinfo {year} {2001})\ pp.\ \bibinfo {pages} {57--66}\BibitemShut {NoStop}%
\bibitem [{\citenamefont {Kempe}\ \emph {et~al.}(2003)\citenamefont {Kempe},
  \citenamefont {Kleinberg},\ and\ \citenamefont {Tardos}}]{Kempe03}%
  \BibitemOpen
  \bibfield  {author} {\bibinfo {author} {\bibfnamefont {D.}~\bibnamefont
  {Kempe}}, \bibinfo {author} {\bibfnamefont {J.}~\bibnamefont {Kleinberg}}, \
  and\ \bibinfo {author} {\bibfnamefont {E.}~\bibnamefont {Tardos}},\ }in\
  \href {\doibase 10.1145/956750.956769} {\emph {\bibinfo {booktitle}
  {Proceedings of the ninth ACM SIGKDD international conference on Knowledge
  discovery and data mining}}},\ \bibinfo {series and number} {KDD '03}\
  (\bibinfo  {publisher} {ACM},\ \bibinfo {address} {New York, NY, USA},\
  \bibinfo {year} {2003})\ pp.\ \bibinfo {pages} {137--146}\BibitemShut
  {NoStop}%
\bibitem [{\citenamefont {Morone}\ and\ \citenamefont
  {Makse}(2015)}]{Morone15}%
  \BibitemOpen
  \bibfield  {author} {\bibinfo {author} {\bibfnamefont {F.}~\bibnamefont
  {Morone}}\ and\ \bibinfo {author} {\bibfnamefont {H.~A.}\ \bibnamefont
  {Makse}},\ }\href {\doibase 10.1038/nature14604} {\bibfield  {journal}
  {\bibinfo  {journal} {Nature}\ }\textbf {\bibinfo {volume} {524}},\ \bibinfo
  {pages} {65} (\bibinfo {year} {2015})}\BibitemShut {NoStop}%
\bibitem [{\citenamefont {Clusella}\ \emph {et~al.}(2016)\citenamefont
  {Clusella}, \citenamefont {Grassberger}, \citenamefont {P\'erez-Reche},\ and\
  \citenamefont {Politi}}]{Clusella16}%
  \BibitemOpen
  \bibfield  {author} {\bibinfo {author} {\bibfnamefont {P.}~\bibnamefont
  {Clusella}}, \bibinfo {author} {\bibfnamefont {P.}~\bibnamefont
  {Grassberger}}, \bibinfo {author} {\bibfnamefont {F.~J.}\ \bibnamefont
  {P\'erez-Reche}}, \ and\ \bibinfo {author} {\bibfnamefont {A.}~\bibnamefont
  {Politi}},\ }\href {\doibase 10.1103/PhysRevLett.117.208301} {\bibfield
  {journal} {\bibinfo  {journal} {Phys. Rev. Lett.}\ }\textbf {\bibinfo
  {volume} {117}},\ \bibinfo {pages} {208301} (\bibinfo {year}
  {2016})}\BibitemShut {NoStop}%
\bibitem [{\citenamefont {Braunstein}\ \emph {et~al.}(2016)\citenamefont
  {Braunstein}, \citenamefont {Dall’Asta}, \citenamefont {Semerjian},\ and\
  \citenamefont {Zdeborová}}]{Braunstein16}%
  \BibitemOpen
  \bibfield  {author} {\bibinfo {author} {\bibfnamefont {A.}~\bibnamefont
  {Braunstein}}, \bibinfo {author} {\bibfnamefont {L.}~\bibnamefont
  {Dall’Asta}}, \bibinfo {author} {\bibfnamefont {G.}~\bibnamefont
  {Semerjian}}, \ and\ \bibinfo {author} {\bibfnamefont {L.}~\bibnamefont
  {Zdeborová}},\ }\href {\doibase 10.1073/pnas.1605083113} {\bibfield
  {journal} {\bibinfo  {journal} {Proceedings of the National Academy of
  Sciences}\ }\textbf {\bibinfo {volume} {113}},\ \bibinfo {pages} {12368}
  (\bibinfo {year} {2016})}\BibitemShut {NoStop}%
\bibitem [{\citenamefont {Mugisha}\ and\ \citenamefont
  {Zhou}(2016)}]{Mugisha16}%
  \BibitemOpen
  \bibfield  {author} {\bibinfo {author} {\bibfnamefont {S.}~\bibnamefont
  {Mugisha}}\ and\ \bibinfo {author} {\bibfnamefont {H.-J.}\ \bibnamefont
  {Zhou}},\ }\href {\doibase 10.1103/PhysRevE.94.012305} {\bibfield  {journal}
  {\bibinfo  {journal} {Phys. Rev. E}\ }\textbf {\bibinfo {volume} {94}},\
  \bibinfo {pages} {012305} (\bibinfo {year} {2016})}\BibitemShut {NoStop}%
\bibitem [{\citenamefont {Zdeborov{\'a}}\ \emph {et~al.}(2016)\citenamefont
  {Zdeborov{\'a}}, \citenamefont {Zhang},\ and\ \citenamefont
  {Zhou}}]{Zdeborova16}%
  \BibitemOpen
  \bibfield  {author} {\bibinfo {author} {\bibfnamefont {L.}~\bibnamefont
  {Zdeborov{\'a}}}, \bibinfo {author} {\bibfnamefont {P.}~\bibnamefont
  {Zhang}}, \ and\ \bibinfo {author} {\bibfnamefont {H.-J.}\ \bibnamefont
  {Zhou}},\ }\href {\doibase 10.1038/srep37954} {\bibfield  {journal} {\bibinfo
   {journal} {Scientific Reports}\ }\textbf {\bibinfo {volume} {6}},\ \bibinfo
  {pages} {37954} (\bibinfo {year} {2016})}\BibitemShut {NoStop}%
\bibitem [{\citenamefont {{Pei}}\ \emph {et~al.}(2016)\citenamefont {{Pei}},
  \citenamefont {{Teng}}, \citenamefont {{Shaman}}, \citenamefont {{Morone}},\
  and\ \citenamefont {{Makse}}}]{Pei16}%
  \BibitemOpen
  \bibfield  {author} {\bibinfo {author} {\bibfnamefont {S.}~\bibnamefont
  {{Pei}}}, \bibinfo {author} {\bibfnamefont {X.}~\bibnamefont {{Teng}}},
  \bibinfo {author} {\bibfnamefont {J.}~\bibnamefont {{Shaman}}}, \bibinfo
  {author} {\bibfnamefont {F.}~\bibnamefont {{Morone}}}, \ and\ \bibinfo
  {author} {\bibfnamefont {H.~A.}\ \bibnamefont {{Makse}}},\ }\href@noop {}
  {\bibfield  {journal} {\bibinfo  {journal} {ArXiv e-prints}\ } (\bibinfo
  {year} {2016})},\ \Eprint {http://arxiv.org/abs/1606.02739} {arXiv:1606.02739
  [physics.soc-ph]} \BibitemShut {NoStop}%
\bibitem [{\citenamefont {Liu}\ \emph {et~al.}(2016)\citenamefont {Liu},
  \citenamefont {Tang}, \citenamefont {Zhou},\ and\ \citenamefont
  {Do}}]{Liu16}%
  \BibitemOpen
  \bibfield  {author} {\bibinfo {author} {\bibfnamefont {Y.}~\bibnamefont
  {Liu}}, \bibinfo {author} {\bibfnamefont {M.}~\bibnamefont {Tang}}, \bibinfo
  {author} {\bibfnamefont {T.}~\bibnamefont {Zhou}}, \ and\ \bibinfo {author}
  {\bibfnamefont {Y.}~\bibnamefont {Do}},\ }\href {\doibase
  http://dx.doi.org/10.1016/j.physa.2016.02.028} {\bibfield  {journal}
  {\bibinfo  {journal} {Physica A: Statistical Mechanics and its Applications}\
  }\textbf {\bibinfo {volume} {452}},\ \bibinfo {pages} {289 } (\bibinfo {year}
  {2016})}\BibitemShut {NoStop}%
\bibitem [{\citenamefont {Szolnoki}\ and\ \citenamefont
  {Perc}(2016)}]{Szolnoki16}%
  \BibitemOpen
  \bibfield  {author} {\bibinfo {author} {\bibfnamefont {A.}~\bibnamefont
  {Szolnoki}}\ and\ \bibinfo {author} {\bibfnamefont {M.}~\bibnamefont
  {Perc}},\ }\href {http://stacks.iop.org/0295-5075/113/i=5/a=58004} {\bibfield
   {journal} {\bibinfo  {journal} {EPL (Europhysics Letters)}\ }\textbf
  {\bibinfo {volume} {113}},\ \bibinfo {pages} {58004} (\bibinfo {year}
  {2016})}\BibitemShut {NoStop}%
\bibitem [{\citenamefont {Gong}\ \emph {et~al.}(2016)\citenamefont {Gong},
  \citenamefont {Yan}, \citenamefont {Shen}, \citenamefont {Ma},\ and\
  \citenamefont {Cai}}]{Gong16}%
  \BibitemOpen
  \bibfield  {author} {\bibinfo {author} {\bibfnamefont {M.}~\bibnamefont
  {Gong}}, \bibinfo {author} {\bibfnamefont {J.}~\bibnamefont {Yan}}, \bibinfo
  {author} {\bibfnamefont {B.}~\bibnamefont {Shen}}, \bibinfo {author}
  {\bibfnamefont {L.}~\bibnamefont {Ma}}, \ and\ \bibinfo {author}
  {\bibfnamefont {Q.}~\bibnamefont {Cai}},\ }\href {\doibase
  http://dx.doi.org/10.1016/j.ins.2016.07.012} {\bibfield  {journal} {\bibinfo
  {journal} {Information Sciences}\ }\textbf {\bibinfo {volume} {367 - 368}},\
  \bibinfo {pages} {600 } (\bibinfo {year} {2016})}\BibitemShut {NoStop}%
\bibitem [{\citenamefont {Wang}\ \emph {et~al.}(2016)\citenamefont {Wang},
  \citenamefont {Zhao}, \citenamefont {Xi},\ and\ \citenamefont {Du}}]{Wang16}%
  \BibitemOpen
  \bibfield  {author} {\bibinfo {author} {\bibfnamefont {Z.}~\bibnamefont
  {Wang}}, \bibinfo {author} {\bibfnamefont {Y.}~\bibnamefont {Zhao}}, \bibinfo
  {author} {\bibfnamefont {J.}~\bibnamefont {Xi}}, \ and\ \bibinfo {author}
  {\bibfnamefont {C.}~\bibnamefont {Du}},\ }\href {\doibase
  http://dx.doi.org/10.1016/j.physa.2016.05.048} {\bibfield  {journal}
  {\bibinfo  {journal} {Physica A: Statistical Mechanics and its Applications}\
  }\textbf {\bibinfo {volume} {461}},\ \bibinfo {pages} {171 } (\bibinfo {year}
  {2016})}\BibitemShut {NoStop}%
\bibitem [{\citenamefont {{Lokhov}}\ and\ \citenamefont
  {{Saad}}(2016)}]{lokhov2016}%
  \BibitemOpen
  \bibfield  {author} {\bibinfo {author} {\bibfnamefont {A.~Y.}\ \bibnamefont
  {{Lokhov}}}\ and\ \bibinfo {author} {\bibfnamefont {D.}~\bibnamefont
  {{Saad}}},\ }\href@noop {} {\bibfield  {journal} {\bibinfo  {journal} {ArXiv
  e-prints}\ } (\bibinfo {year} {2016})},\ \Eprint
  {http://arxiv.org/abs/1608.08278} {arXiv:1608.08278} \BibitemShut {NoStop}%
\bibitem [{\citenamefont {Ni}(2016)}]{Ni16}%
  \BibitemOpen
  \bibfield  {author} {\bibinfo {author} {\bibfnamefont {Y.}~\bibnamefont
  {Ni}},\ }\href {\doibase 10.1016/j.asoc.2016.04.025} {\bibfield  {journal}
  {\bibinfo  {journal} {Applied Soft Computing}\ ,\ } (\bibinfo {year}
  {2016})}\BibitemShut {NoStop}%
\bibitem [{\citenamefont {{Cordasco}}\ \emph {et~al.}(2015)\citenamefont
  {{Cordasco}}, \citenamefont {{Gargano}}, \citenamefont {{Rescigno}},\ and\
  \citenamefont {{Vaccaro}}}]{cordasco2015}%
  \BibitemOpen
  \bibfield  {author} {\bibinfo {author} {\bibfnamefont {G.}~\bibnamefont
  {{Cordasco}}}, \bibinfo {author} {\bibfnamefont {L.}~\bibnamefont
  {{Gargano}}}, \bibinfo {author} {\bibfnamefont {A.~A.}\ \bibnamefont
  {{Rescigno}}}, \ and\ \bibinfo {author} {\bibfnamefont {U.}~\bibnamefont
  {{Vaccaro}}},\ }\href@noop {} {\bibfield  {journal} {\bibinfo  {journal}
  {ArXiv e-prints}\ } (\bibinfo {year} {2015})},\ \Eprint
  {http://arxiv.org/abs/1512.06372} {arXiv:1512.06372 [cs.DS]} \BibitemShut
  {NoStop}%
\bibitem [{\citenamefont {Guo}\ \emph {et~al.}(2016)\citenamefont {Guo},
  \citenamefont {Lin}, \citenamefont {Guo},\ and\ \citenamefont {Liu}}]{Guo16}%
  \BibitemOpen
  \bibfield  {author} {\bibinfo {author} {\bibfnamefont {L.}~\bibnamefont
  {Guo}}, \bibinfo {author} {\bibfnamefont {J.-H.}\ \bibnamefont {Lin}},
  \bibinfo {author} {\bibfnamefont {Q.}~\bibnamefont {Guo}}, \ and\ \bibinfo
  {author} {\bibfnamefont {J.-G.}\ \bibnamefont {Liu}},\ }\href {\doibase
  http://dx.doi.org/10.1016/j.physleta.2015.12.031} {\bibfield  {journal}
  {\bibinfo  {journal} {Physics Letters A}\ }\textbf {\bibinfo {volume}
  {380}},\ \bibinfo {pages} {837 } (\bibinfo {year} {2016})}\BibitemShut
  {NoStop}%
\bibitem [{\citenamefont {{Gao}}\ \emph {et~al.}(2016)\citenamefont {{Gao}},
  \citenamefont {{Wang}}, \citenamefont {{Pan}}, \citenamefont {{Tang}},\ and\
  \citenamefont {{Zhang}}}]{gao2016}%
  \BibitemOpen
  \bibfield  {author} {\bibinfo {author} {\bibfnamefont {L.}~\bibnamefont
  {{Gao}}}, \bibinfo {author} {\bibfnamefont {W.}~\bibnamefont {{Wang}}},
  \bibinfo {author} {\bibfnamefont {L.}~\bibnamefont {{Pan}}}, \bibinfo
  {author} {\bibfnamefont {M.}~\bibnamefont {{Tang}}}, \ and\ \bibinfo {author}
  {\bibfnamefont {H.-F.}\ \bibnamefont {{Zhang}}},\ }\href {\doibase
  10.1038/srep38220} {\bibfield  {journal} {\bibinfo  {journal} {Scientific
  Reports}\ }\textbf {\bibinfo {volume} {6}},\ \bibinfo {eid} {38220} (\bibinfo
  {year} {2016})},\ \Eprint {http://arxiv.org/abs/1606.05408} {arXiv:1606.05408
  [physics.soc-ph]} \BibitemShut {NoStop}%
\bibitem [{\citenamefont {Achlioptas}\ \emph {et~al.}(2009)\citenamefont
  {Achlioptas}, \citenamefont {D{\textquoteright}Souza},\ and\ \citenamefont
  {Spencer}}]{achlioptas2009explosive}%
  \BibitemOpen
  \bibfield  {author} {\bibinfo {author} {\bibfnamefont {D.}~\bibnamefont
  {Achlioptas}}, \bibinfo {author} {\bibfnamefont {R.~M.}\ \bibnamefont
  {D{\textquoteright}Souza}}, \ and\ \bibinfo {author} {\bibfnamefont
  {J.}~\bibnamefont {Spencer}},\ }\href {\doibase 10.1126/science.1167782}
  {\bibfield  {journal} {\bibinfo  {journal} {Science}\ }\textbf {\bibinfo
  {volume} {323}},\ \bibinfo {pages} {1453} (\bibinfo {year}
  {2009})}\BibitemShut {NoStop}%
\bibitem [{\citenamefont {D'Souza}\ and\ \citenamefont
  {Nagler}(2015)}]{d2015anomalous}%
  \BibitemOpen
  \bibfield  {author} {\bibinfo {author} {\bibfnamefont {R.~M.}\ \bibnamefont
  {D'Souza}}\ and\ \bibinfo {author} {\bibfnamefont {J.}~\bibnamefont
  {Nagler}},\ }\href {\doibase 10.1038/nphys3378} {\bibfield  {journal}
  {\bibinfo  {journal} {Nature Physics}\ }\textbf {\bibinfo {volume} {11}},\
  \bibinfo {pages} {531} (\bibinfo {year} {2015})}\BibitemShut {NoStop}%
\bibitem [{\citenamefont {Goldenberg}\ \emph {et~al.}(2001)\citenamefont
  {Goldenberg}, \citenamefont {Libai},\ and\ \citenamefont
  {Muller}}]{Goldenberg01}%
  \BibitemOpen
  \bibfield  {author} {\bibinfo {author} {\bibfnamefont {J.}~\bibnamefont
  {Goldenberg}}, \bibinfo {author} {\bibfnamefont {B.}~\bibnamefont {Libai}}, \
  and\ \bibinfo {author} {\bibfnamefont {E.}~\bibnamefont {Muller}},\ }\href
  {\doibase 10.1023/A:1011122126881} {\bibfield  {journal} {\bibinfo  {journal}
  {Marketing Letters}\ }\textbf {\bibinfo {volume} {12}},\ \bibinfo {pages}
  {211} (\bibinfo {year} {2001})}\BibitemShut {NoStop}%
\bibitem [{\citenamefont {Leskovec}\ \emph {et~al.}(2007)\citenamefont
  {Leskovec}, \citenamefont {Krause}, \citenamefont {Guestrin}, \citenamefont
  {Faloutsos}, \citenamefont {VanBriesen},\ and\ \citenamefont
  {Glance}}]{Leskovec07}%
  \BibitemOpen
  \bibfield  {author} {\bibinfo {author} {\bibfnamefont {J.}~\bibnamefont
  {Leskovec}}, \bibinfo {author} {\bibfnamefont {A.}~\bibnamefont {Krause}},
  \bibinfo {author} {\bibfnamefont {C.}~\bibnamefont {Guestrin}}, \bibinfo
  {author} {\bibfnamefont {C.}~\bibnamefont {Faloutsos}}, \bibinfo {author}
  {\bibfnamefont {J.}~\bibnamefont {VanBriesen}}, \ and\ \bibinfo {author}
  {\bibfnamefont {N.}~\bibnamefont {Glance}},\ }in\ \href {\doibase
  10.1145/1281192.1281239} {\emph {\bibinfo {booktitle} {Proceedings of the
  13th ACM SIGKDD International Conference on Knowledge Discovery and Data
  Mining}}},\ \bibinfo {series and number} {KDD '07}\ (\bibinfo  {publisher}
  {ACM},\ \bibinfo {address} {New York, NY, USA},\ \bibinfo {year} {2007})\
  pp.\ \bibinfo {pages} {420--429}\BibitemShut {NoStop}%
\bibitem [{\citenamefont {Chen}\ \emph {et~al.}(2009)\citenamefont {Chen},
  \citenamefont {Wang},\ and\ \citenamefont {Yang}}]{Chen09}%
  \BibitemOpen
  \bibfield  {author} {\bibinfo {author} {\bibfnamefont {W.}~\bibnamefont
  {Chen}}, \bibinfo {author} {\bibfnamefont {Y.}~\bibnamefont {Wang}}, \ and\
  \bibinfo {author} {\bibfnamefont {S.}~\bibnamefont {Yang}},\ }in\ \href
  {\doibase 10.1145/1557019.1557047} {\emph {\bibinfo {booktitle} {Proceedings
  of the 15th ACM SIGKDD International Conference on Knowledge Discovery and
  Data Mining}}},\ \bibinfo {series and number} {KDD '09}\ (\bibinfo
  {publisher} {ACM},\ \bibinfo {address} {New York, NY, USA},\ \bibinfo {year}
  {2009})\ pp.\ \bibinfo {pages} {199--208}\BibitemShut {NoStop}%
\bibitem [{\citenamefont {{Nguyen}}\ \emph {et~al.}(2016)\citenamefont
  {{Nguyen}}, \citenamefont {{Thai}},\ and\ \citenamefont {{Dinh}}}]{Nguyen16}%
  \BibitemOpen
  \bibfield  {author} {\bibinfo {author} {\bibfnamefont {H.~T.}\ \bibnamefont
  {{Nguyen}}}, \bibinfo {author} {\bibfnamefont {M.~T.}\ \bibnamefont
  {{Thai}}}, \ and\ \bibinfo {author} {\bibfnamefont {T.~N.}\ \bibnamefont
  {{Dinh}}},\ }\href@noop {} {\bibfield  {journal} {\bibinfo  {journal} {ArXiv
  e-prints}\ } (\bibinfo {year} {2016})},\ \Eprint
  {http://arxiv.org/abs/1605.07990} {arXiv:1605.07990} \BibitemShut {NoStop}%
\bibitem [{\citenamefont {Altarelli}\ \emph {et~al.}(2013)\citenamefont
  {Altarelli}, \citenamefont {Braunstein}, \citenamefont {Dall’Asta},\ and\
  \citenamefont {Zecchina}}]{Altarelli13}%
  \BibitemOpen
  \bibfield  {author} {\bibinfo {author} {\bibfnamefont {F.}~\bibnamefont
  {Altarelli}}, \bibinfo {author} {\bibfnamefont {A.}~\bibnamefont
  {Braunstein}}, \bibinfo {author} {\bibfnamefont {L.}~\bibnamefont
  {Dall’Asta}}, \ and\ \bibinfo {author} {\bibfnamefont {R.}~\bibnamefont
  {Zecchina}},\ }\href {http://stacks.iop.org/1742-5468/2013/i=09/a=P09011}
  {\bibfield  {journal} {\bibinfo  {journal} {Journal of Statistical Mechanics:
  Theory and Experiment}\ }\textbf {\bibinfo {volume} {2013}},\ \bibinfo
  {pages} {P09011} (\bibinfo {year} {2013})}\BibitemShut {NoStop}%
\bibitem [{\citenamefont {Jian-Xiong~Zhang}\ and\ \citenamefont
  {Zhao}(2016)}]{Zhang16}%
  \BibitemOpen
  \bibfield  {author} {\bibinfo {author} {\bibfnamefont {Q.~D.}\ \bibnamefont
  {Jian-Xiong~Zhang}, \bibfnamefont {Duan-Bing~Chen}}\ and\ \bibinfo {author}
  {\bibfnamefont {Z.-D.}\ \bibnamefont {Zhao}},\ }\href {\doibase
  10.1038/srep27823} {\bibfield  {journal} {\bibinfo  {journal} {Scientific
  Reports}\ }\textbf {\bibinfo {volume} {6}},\ \bibinfo {pages} {27823}
  (\bibinfo {year} {2016})}\BibitemShut {NoStop}%
\bibitem [{\citenamefont {Newman}(2002)}]{Newman02}%
  \BibitemOpen
  \bibfield  {author} {\bibinfo {author} {\bibfnamefont {M.~E.~J.}\
  \bibnamefont {Newman}},\ }\href {\doibase 10.1103/PhysRevE.66.016128}
  {\bibfield  {journal} {\bibinfo  {journal} {Phys. Rev. E}\ }\textbf {\bibinfo
  {volume} {66}},\ \bibinfo {pages} {016128} (\bibinfo {year}
  {2002})}\BibitemShut {NoStop}%
\bibitem [{\citenamefont {Kitsak}\ \emph {et~al.}(2010)\citenamefont {Kitsak},
  \citenamefont {Gallos}, \citenamefont {Havlin}, \citenamefont {Liljeros},
  \citenamefont {Muchnik}, \citenamefont {Stanley},\ and\ \citenamefont
  {Makse}}]{Kitsak10}%
  \BibitemOpen
  \bibfield  {author} {\bibinfo {author} {\bibfnamefont {M.}~\bibnamefont
  {Kitsak}}, \bibinfo {author} {\bibfnamefont {L.}~\bibnamefont {Gallos}},
  \bibinfo {author} {\bibfnamefont {S.}~\bibnamefont {Havlin}}, \bibinfo
  {author} {\bibfnamefont {F.}~\bibnamefont {Liljeros}}, \bibinfo {author}
  {\bibfnamefont {L.}~\bibnamefont {Muchnik}}, \bibinfo {author} {\bibfnamefont
  {H.}~\bibnamefont {Stanley}}, \ and\ \bibinfo {author} {\bibfnamefont
  {H.}~\bibnamefont {Makse}},\ }\href {\doibase 10.1038/nphys1746} {\bibfield
  {journal} {\bibinfo  {journal} {Nature Physics}\ }\textbf {\bibinfo {volume}
  {6}},\ \bibinfo {pages} {888} (\bibinfo {year} {2010})}\BibitemShut {NoStop}%
\bibitem [{\citenamefont {Radicchi}\ and\ \citenamefont
  {Castellano}(2016)}]{Radicchi16}%
  \BibitemOpen
  \bibfield  {author} {\bibinfo {author} {\bibfnamefont {F.}~\bibnamefont
  {Radicchi}}\ and\ \bibinfo {author} {\bibfnamefont {C.}~\bibnamefont
  {Castellano}},\ }\href {\doibase 10.1103/PhysRevE.93.062314} {\bibfield
  {journal} {\bibinfo  {journal} {Phys. Rev. E}\ }\textbf {\bibinfo {volume}
  {93}},\ \bibinfo {pages} {062314} (\bibinfo {year} {2016})}\BibitemShut
  {NoStop}%
\bibitem {SM}%
  \BibitemOpen
  \bibinfo {note} {See Supplemental Material at [URL will be inserted by
  publisher] for results for $p \not =p_c$ and details on the networks
  considered.}\BibitemShut {Stop}%
\bibitem [{\citenamefont {Castellano}\ and\ \citenamefont
  {Pastor-Satorras}(2016)}]{Castellano16}%
  \BibitemOpen
  \bibfield  {author} {\bibinfo {author} {\bibfnamefont {C.}~\bibnamefont
  {Castellano}}\ and\ \bibinfo {author} {\bibfnamefont {R.}~\bibnamefont
  {Pastor-Satorras}},\ }\href {\doibase 10.1140/epjb/e2016-60953-5} {\bibfield
  {journal} {\bibinfo  {journal} {The European Physical Journal B}\ }\textbf
  {\bibinfo {volume} {89}},\ \bibinfo {pages} {243} (\bibinfo {year}
  {2016})}\BibitemShut {NoStop}%
\end{thebibliography}

%

\newpage
\pagestyle{empty}





\section*{Supplementary material}

# Supplemental Material: "Fundamental difference between superblockers and superspreaders in networks"

Filippo Radicchi
*Center for Complex Networks and Systems Research,*
*School of Informatics and Computing, Indiana University, Bloomington, USA\**

Claudio Castellano
*Istituto dei Sistemi Complessi (ISC-CNR), Roma, Italy,*
*and Dipartimento di Fisica, Sapienza Università di Roma, Roma, Italy*

---

\* filiradi@indiana.edu

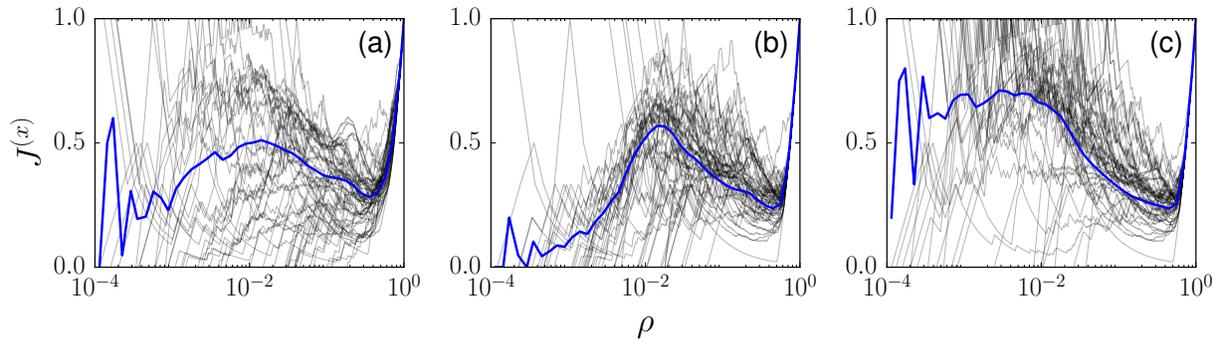

Figure SM1. Same as Fig. 1 of the main text but for $p = p_c/2$.

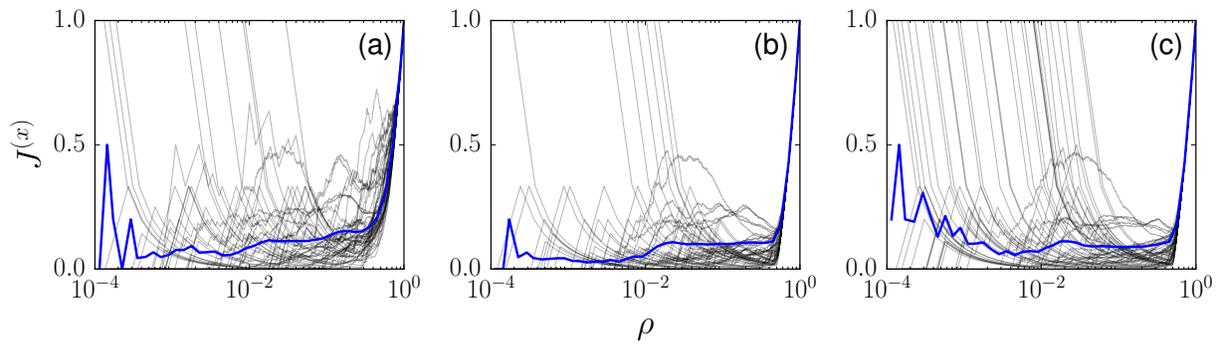

Figure SM2. Same as Fig. 1 of the main text but for $p = 2\,p_c$.

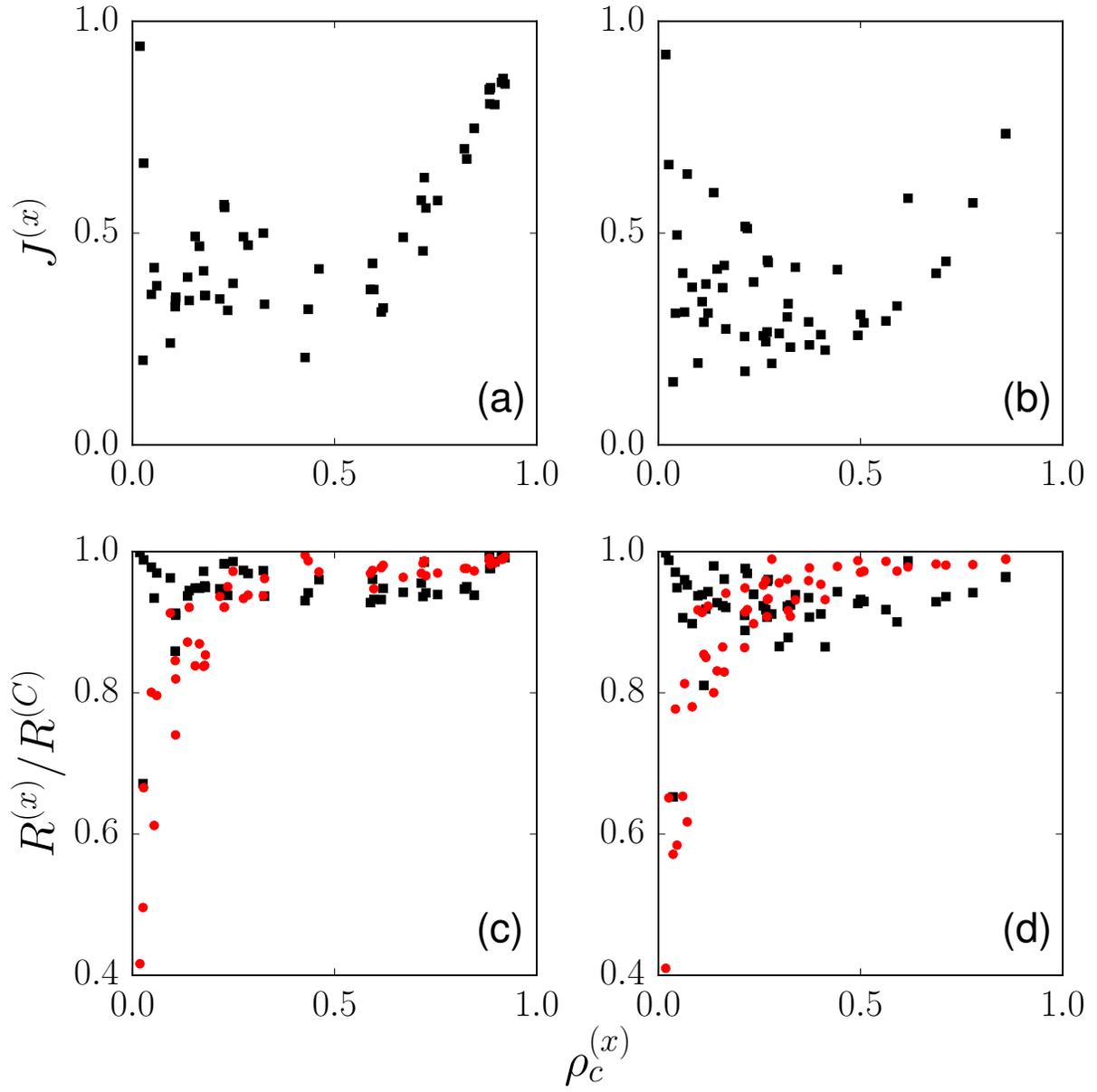

Figure SM3. Same as Fig. 2 of the main text but for $p = p_c/2$.

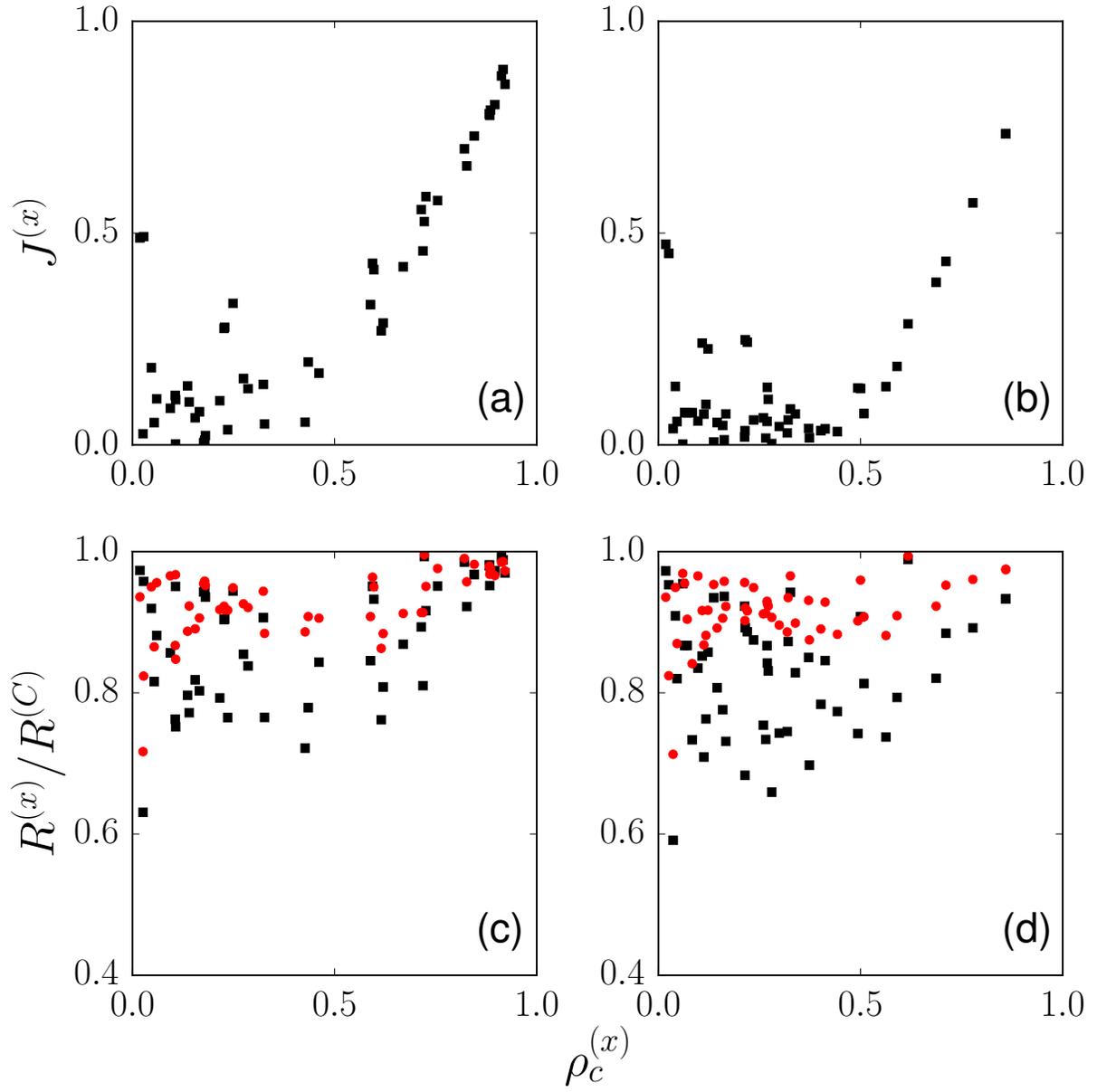

Figure SM4. Same as Fig. 2 of the main text but for $p = 2\,p_c$.

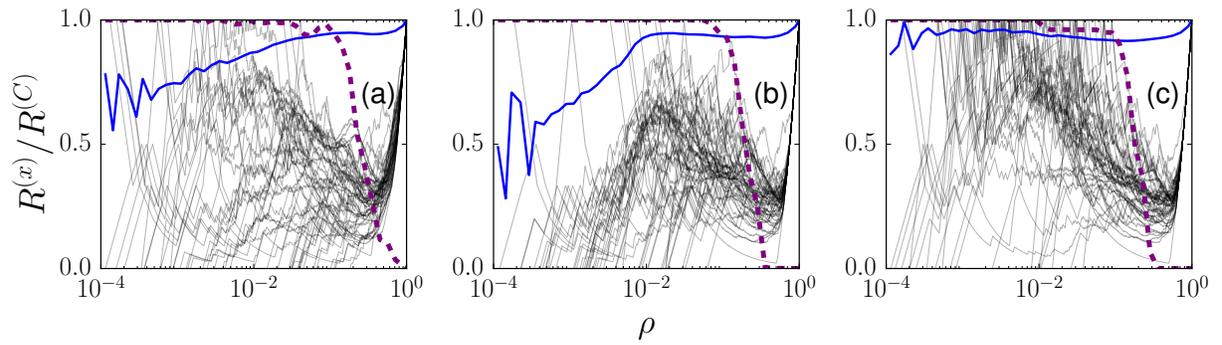

Figure SM5. Same as Fig. 3 of the main text but for $p = p_c/2$.

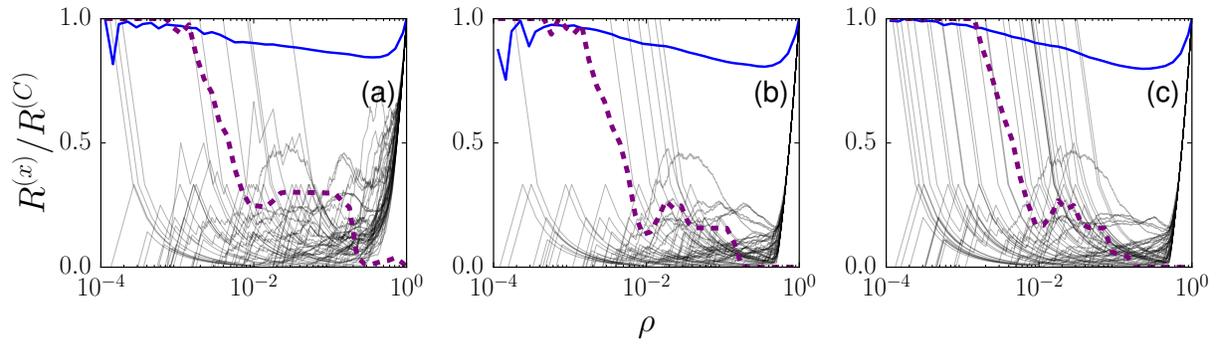

Figure SM6. Same as Fig. 3 of the main text but for $p = 2\,p_c$.

| rank | network | $N$ | $p_c$ | Refs. | Url |
|---|---|---|---|---|---|
| 1 | Social 3 | 32 | 0.180 | [1] | url |
| 2 | Karate club | 34 | 0.170 | [2] | url |
| 3 | Protein 2 | 53 | 0.220 | [1] | url |
| 4 | Dolphins | 62 | 0.170 | [3] | url |
| 5 | Social 1 | 67 | 0.230 | [1] | url |
| 6 | Les Miserables | 77 | 0.110 | [4] | url |
| 7 | Protein 1 | 95 | 0.260 | [1] | url |
| 8 | E. Coli, transcription | 97 | 0.230 | [5] | url |
| 9 | Political books | 105 | 0.100 | [6] | url |
| 10 | David Copperfield | 112 | 0.085 | [7] | url |
| 11 | College football | 115 | 0.105 | [8] | url |
| 12 | S 208 | 122 | 0.350 | [1] | url |
| 13 | High school, 2011 | 126 | 0.030 | [9] | url |
| 14 | Bay Dry | 128 | 0.025 | [10, 11] | url |
| 15 | Bay Wet | 128 | 0.025 | [11] | url |
| 16 | Radoslaw Email | 167 | 0.015 | [11, 12] | url |
| 17 | High school, 2012 | 180 | 0.035 | [9] | url |
| 18 | Little Rock Lake | 183 | 0.025 | [11, 13] | url |
| 19 | Jazz | 198 | 0.025 | [14] | url |
| 20 | S 420 | 252 | 0.360 | [1] | url |
| 21 | C. Elegans, neural | 297 | 0.045 | [15] | url |
| 22 | Dublin | 410 | 0.060 | [11, 16] | url |
| 23 | US Air Trasportation | 500 | 0.026 | [17] | url |
| 24 | S 838 | 512 | 0.349 | [1] | url |
| 25 | Yeast, transcription | 688 | 0.192 | [18] | url |
| 26 | URV email | 1,133 | 0.056 | [19] | url |

Table SM1. Real-world networks included in our analysis. The index appearing on the leftmost column serves only as a counter for the network analyzed. The following columns report: the name of the network, the number of nodes in the network, the best estimate of the critical value of the probability for the ICM, reference(s) of the paper where the network has been first analyzed, and url of where the network data have been downloaded (to open the web page in your browser, just click on the word "url").

| rank | network | $N$ | $p_c$ | Refs. | Url |
|------|---------|-----|-------|-------|-----|
| 27 | Political blogs | $1,224$ | 0.015 | [6] | url |
| 28 | Air traffic | $1,226$ | 0.163 | [11] | url |
| 29 | Network Science | $1,461$ | 0.323 | [7] | url |
| 30 | Yeast, protein | $1,846$ | 0.253 | [20] | url |
| 31 | Petster, hamster | $1,858$ | 0.025 | [11] | url |
| 32 | UC Irvine | $1,899$ | 0.023 | [11, 21] | url |
| 33 | Yeast, protein | $2,284$ | 0.071 | [22] | url |
| 34 | Japanese | $2,704$ | 0.030 | [1] | url |
| 35 | Open flights | $2,939$ | 0.020 | [11, 23] | url |
| 36 | Tennis | $4,342$ | 0.007 | [24] | url |
| 37 | US Power grid | $4,941$ | 0.437 | [15] | url |
| 38 | GR-QC, 1993-2003 | $5,241$ | 0.091 | [25] | url |
| 39 | HT09 | $5,352$ | 0.025 | [16] | url |
| 40 | Jung | $6,120$ | 0.008 | [11, 26] | url |
| 41 | Reactome | $6,229$ | 0.006 | [11, 27] | url |
| 42 | Gnutella, Aug. 8, 2002 | $6,301$ | 0.048 | [25, 28] | url |
| 43 | JDK | $6,434$ | 0.008 | [11] | url |
| 44 | AS Oregon | $6,474$ | 0.033 | [29] | url |
| 45 | English | $7,381$ | 0.011 | [1] | url |
| 46 | Hep-Th, 1995-1999 | $7,610$ | 0.113 | [30] | url |
| 47 | Gnutella, Aug. 9, 2002 | $8,114$ | 0.049 | [25, 28] | url |
| 48 | French | $8,325$ | 0.022 | [1] | url |
| 49 | Gnutella, Aug. 6, 2002 | $8,717$ | 0.062 | [25, 28] | url |
| 50 | Gnutella, Aug. 5, 2002 | $8,846$ | 0.059 | [25, 28] | url |
| 51 | Hep-Th, 1993-2003 | $9,875$ | 0.072 | [25] | url |

Table SM2. Continuation of Table SM1.

[1] R. Milo, S. Itzkovitz, N. Kashtan, R. Levitt, S. Shen-Orr, I. Ayzenshtat, M. Sheffer, and U. Alon, Science **303**, 1538 (2004).

[2] W. W. Zachary, Journal of anthropological research , 452 (1977).

[3] D. Lusseau, K. Schneider, O. J. Boisseau, P. Haase, E. Slooten, and S. M. Dawson, Behavioral Ecology and Sociobiology **54**, 396 (2003).

[4] D. E. Knuth, D. E. Knuth, and D. E. Knuth, *The Stanford GraphBase: a platform for combinatorial computing*, Vol. 37 (Addison-Wesley Reading, 1993).

[5] S. Mangan and U. Alon, Proceedings of the National Academy of Sciences **100**, 11980 (2003).

[6] L. A. Adamic and N. Glance, in *Proceedings of the 3rd international workshop on Link discovery* (ACM, 2005) pp. 36–43.

[7] M. E. Newman, Physical review E **74**, 036104 (2006).

[8] M. Girvan and M. E. Newman, Proceedings of the National Academy of Sciences **99**, 7821 (2002).

[9] J. Fournet and A. Barrat, PloS one **9**, e107878 (2014).

[10] R. Ulanowicz, C. Bondavalli, and M. Egnotovich, Annual Report to the United States Geological Service Biological Resources Division Ref. No.[UMCES] CBL , 98 (1998).

[11] J. Kunegis, in *Proc. Int. Conf. on World Wide Web Companion* (2013) pp. 1343–1350.

[12] R. Michalski, S. Palus, and P. Kazienko, in *Lecture Notes in Business Information Processing*, Vol. 87 (Springer Berlin Heidelberg, 2011) pp. 197–206.

[13] N. D. Martinez, Ecological Monographs , 367 (1991).

[14] P. M. Gleiser and L. Danon, Advances in complex systems **6**, 565 (2003).

[15] D. J. Watts and S. H. Strogatz, nature **393**, 440 (1998).

[16] L. Isella, J. Stehlé, A. Barrat, C. Cattuto, J.-F. Pinton, and W. Van den Broeck, Journal of theoretical biology **271**, 166 (2011).

[17] V. Colizza, R. Pastor-Satorras, and A. Vespignani, Nature Physics **3**, 276 (2007).

[18] R. Milo, S. Shen-Orr, S. Itzkovitz, N. Kashtan, D. Chklovskii, and U. Alon, Science **298**, 824 (2002).

[19] R. Guimera, L. Danon, A. Diaz-Guilera, F. Giralt, and A. Arenas, Physical review E **68**, 065103 (2003).

[20] H. Jeong, S. P. Mason, A.-L. Barabási, and Z. N. Oltvai, Nature **411**, 41 (2001).

[21] T. Opsahl and P. Panzarasa, Social networks **31**, 155 (2009).

[22] D. Bu, Y. Zhao, L. Cai, H. Xue, X. Zhu, H. Lu, J. Zhang, S. Sun, L. Ling, N. Zhang, *et al.*, Nucleic acids research **31**, 2443 (2003).

[23] T. Opsahl, F. Agneessens, and J. Skvoretz, Social Networks **32**, 245 (2010).

[24] F. Radicchi, PloS one **6**, e17249 (2011).

[25] J. Leskovec, J. Kleinberg, and C. Faloutsos, ACM Transactions on Knowledge Discovery from Data (TKDD) **1**, 2 (2007).

[26] L. Šubelj and M. Bajec, in *Proceedings of the First International Workshop on Software Mining* (ACM, 2012) pp. 9–16.

[27] G. Joshi-Tope, M. Gillespie, I. Vastrik, P. D'Eustachio, E. Schmidt, B. de Bono, B. Jassal, G. Gopinath, G. Wu, L. Matthews, *et al.*, Nucleic acids research **33**, D428 (2005).

[28] M. Ripeanu, I. Foster, and A. Iamnitchi, arXiv preprint cs/0209028 (2002).

[29] J. Leskovec, J. Kleinberg, and C. Faloutsos, in *Proceedings of the eleventh ACM SIGKDD international conference on Knowledge discovery in data mining* (ACM, 2005) pp. 177–187.

[30] M. E. Newman, Proceedings of the National Academy of Sciences **98**, 404 (2001).



\end{document}